# Momentum-scalar coupled turbulence with anomalous momentum and scalar diffusions. Part 1: Without external force and with long-range external force

Wei Zhao (赵伟)

*State Key Laboratory of Photon-Technology in Western China Energy, International Scientific and Technological Cooperation Base of Photoelectric Technology and Functional Materials and Application, Laboratory of Optoelectronic Technology of Shaanxi Province, Institute of Photonics and Photon-technology, Northwest University, Xi'an 710127, China*

**Abstract** We present a theoretical model for momentum–scalar coupled turbulence in which both fields undergo anomalous diffusion, described by fractional biharmonic operators of orders $\gamma/4$ and $\alpha/4$, respectively. Focusing on the long-range external forcing or unforced turbulence, we derive analytical expressions for the kinetic energy spectrum $E_u(k)$, the scalar spectrum $E_s(k)$, the characteristic wavenumbers $k_K = \left(\varepsilon_u^{1/3}/c_u\right)^{1/\left(\gamma-\frac{2}{3}\right)}$ (reciprocal of Kolmogorov scale) and $k_S = \left(\varepsilon_u^{1/3}/c_s\right)^{1/\left(\alpha-\frac{2}{3}\right)}$ (reciprocal of scalar dissipation scale) as functions of $\gamma$, $\alpha$, turbulent dissipation rate $\varepsilon_u$, diffusivities of momentum ($c_u$) and scalar ($c_s$), respectively. An anomalous Schmidt number $Sc_Z = k_0^{\gamma-\alpha} c_u/c_s$ is defined to governs the cascade topology. It describes the ratio of diffusion times of scalar and momentum on the minimum wavenumber $k_0$. Superdiffusion ($\gamma < 2$ or $\alpha < 2$) is shown to counter-intuitively enlarge $k_K$ and $k_S$, broadening the inertial range. The theory unifies the classical Kolmogorov–Obukhov–Corrsin–Batchelor scalings as special cases when $\gamma = \alpha = 2$, and provides a foundation for understanding non-Fickian transport in complex turbulent systems.

## 1. Introduction

In turbulence, nonlinear dynamics plays an essential role in the evolution of turbulent kinetic energy and scalar variance as they cascade from large to small scales, where diffusion become important. In classical studies, e.g. hydrodynamic and aerodynamic turbulence, only ordinary (Fickian) diffusion is taken into account. However, in nature and engineering, ordinary diffusion may not always be valid, while anomalous diffusion plays important roles in broad areas. For example, in geophysical fluid dynamics, passive scalar can experience anomalous diffusion in vortical flow [1,2]. In both astrophysical and laboratory plasmas, anomalous diffusions that different from standard Brownian diffusion have been widely investigated as well [3]. In the realm of soft condensed matter and complex fluids, anomalous diffusion is equally pervasive. Many disordered fluid systems exhibit transport dynamics that deviate fundamentally from Brownian motion; examples include the motion of tracers in glassy materials [4], active matter [5] and bacterial flow [6], where crowding and heterogeneous environments induce subdiffusive or superdiffusive behavior depending on the time and length scales of observation. The transport of membrane-bound proteins and lipids in living cells similarly displays anomalous characteristics, reflecting the combined effects of molecular crowding, active cellular, and the viscoelastic nature of the cytoskeleton [7,8]. Even in apparently simple systems such as porous media and amorphous semiconductors, the interplay of geometrical disorder and long-range correlations gives rise to subdiffusion and superdiffusion [9,10]. This broad and growing body of evidence has led to the now-common refrain in the literature that "anomalous is normal"—a recognition that non-Fickian transport is not an exotic exception but rather a generic feature of complex, multi-scale, and multi-physics systems. Anomalous diffusion can physically modify the transports of momentum and scalar at small scales [11], accordingly, change the characteristic length scales of turbulence and their eventual fate—dissipations.

The mathematical modeling of anomalous diffusion has increasingly relied on fractional-order differential equations [12], which possess the inherent ability to describe large-scale behavior with greater efficiency than fully resolved classical models by encoding long-range interactions and long-term memory effects through non-local operators [12]. Many attempts [13-15] have been made focusing on the existence of solution/weak solution of fractional Navier-Stokes equation, regarding a fractional Laplacian operator $(-\Delta)^{\chi}$ of the order $\chi$. In contrast, little efforts have been made to elucidate turbulent structures affected by anomalous diffusion, except for Akhavan-Safaei and Zayernouri [16] who suggest a scalar spectrum theory for scalar turbulence regarding non-local scalar transfer in spectral space denoted via fractional Laplacian operator.

Some efforts have been placed on using fractional derivative to depict external volume force and analyzed its influence on the physical structures of turbulence. For instance, in 2021, Zhao and Wang [17] employed a fractional derivative to characterize the relationship between external volume force and a control scalar. The cascade of the turbulence driven by the volume force have been theoretically analyzed based on the order of derivation. However, the fractional derivative is more in a phenomenology form, not strictly mathematical (due to the incompleteness of fractional calculus in the current stage, similar issue commonly exists in broad applications). Even though, the investigation found some representative orders of derivation, e.g. 2/3 and 3/2 which are corresponding to the 1/3 and 3/4 by Boutros and Gibbon [15], that could affect the cascade of turbulence. Later, by phenomenology analysis on the volume force with the fractional derivative, a general flux model, i.e. Quad-cascade process model [18] was established. To be more rigorous from the aspect of mathematics, the model was further depicted using a fractional biharmonic operator [19]. However, in the investigation, they only considered the influence of ordinary diffusions of momentum and scalars.

To the best of our knowledge, none of the studies simultaneously accounts for anomalous diffusion in both momentum and scalar fields under multiscale forcing—a gap that the present investigation aims to fill. In this investigation, a theoretical model regarding the influence of different diffusions of momentum and scalar has been established via fractional derivations, for momentum-scalar coupled turbulence. In this part, we only discuss the cases that no external forcing or under long-range external force. Particular attention is paid to the scaling exponents in different cascade subranges and the characteristic length scales, revealing how anomalous diffusion reshapes the structure evolution of turbulent flow fields.

The paper is organized as follows. In section 2, we define the fractional biharmonic operator and establish the conservation model governing the momentum–scalar coupled turbulence, for both no external forcing and with long-range external forcing cases. In section 3, we present the analytical solutions for the kinetic energy spectrum and its characteristic wavenumbers. In sections 4 and 5, we derive the scalar spectra and characteristic scales for $Sc_Z \gg 1$ and $Sc_Z \ll 1$ respectively. A discussion of the physical implications and the limitations of the model is provided in section 6, with conclusions in section 7.

## 2. Theory

In this study, fractional biharmonic operator has been employed to comprehensively describe the anomalous diffusion and the external force. Thus, the fractional biharmonic operator is introduced first.

*2.1 Fractional biharmonic operator*

For any function $f: \mathbb{R}^{\mathrm{d}} \to \mathbb{R}$ (where $d$ is the spatial dimension) that is smooth, a fractional biharmonic operator of order $\chi$ can be defined as

$$\mathfrak{D}^{\chi} f = \int_{-\infty}^{+\infty} (k^4)^{\chi} \hat{f}(\boldsymbol{k}, t) e^{i\boldsymbol{k}\cdot\boldsymbol{x}} \mathrm{d}\boldsymbol{k} \tag{1}$$

where $f(\boldsymbol{x}, t) = \int_{-\infty}^{+\infty} \hat{f}(\boldsymbol{k}, t) e^{i\boldsymbol{k}\cdot\boldsymbol{x}} \mathrm{d}\boldsymbol{k}$ and $k = |\boldsymbol{k}|$. $\chi$ can be an integer or a fraction number, thus $(k^4)^{\chi}$ can be complex. For instance, when $\chi = 1/2$,

$$(k^4)^{\frac{1}{2}} = \begin{cases} k^2 \\ -k^2 \end{cases} \text{ and } \mathfrak{D}^{\frac{1}{2}} = \begin{cases} (-\Delta)^1 \\ -(-\Delta)^1 \end{cases} \tag{2}$$

where $(-\Delta)^{2\chi}$ is a fractional Laplacian operator [15,20-22] of order $2\chi$, defined as $(-\Delta)^{2\chi} f = \int_{-\infty}^{+\infty} k^{4\chi} \hat{f}(\boldsymbol{k}, t) e^{i\boldsymbol{k}\cdot\boldsymbol{x}} \mathrm{d}\boldsymbol{k}$. In this case, two subsets can be obtained. When $\chi = 1/4$,

$$(k^4)^{\frac{1}{4}} = \begin{cases} k \\ ik \\ -k \\ -ik \end{cases} \text{ and } \mathfrak{D}^{\frac{1}{4}} = \begin{cases} (-\Delta)^{\frac{1}{2}} \\ i(-\Delta)^{\frac{1}{2}} \\ -(-\Delta)^{\frac{1}{2}} \\ -i(-\Delta)^{\frac{1}{2}} \end{cases} \tag{3}$$

Four subsets are obtained. Thereby, the fractional biharmonic operator provides us a powerful tool with more availability to represent various forms of $\Delta$. What we need to do is select a subset of fractional biharmonic operator for different physical model.

Besides, it can be proved the fractional biharmonic operator still follows

$$\mathfrak{D}^{\chi_1} \mathfrak{D}^{\chi_2} f = \mathfrak{D}^{\chi_1 + \chi_2} f \tag{4}$$

In this investigation, the first and second subsets are selected. The first one is defined as

$$\mathfrak{D}_{(1)}^{\chi} f = \int_{-\infty}^{+\infty} k^{4\chi} \hat{f}(\boldsymbol{k}, t) e^{i\boldsymbol{k}\cdot\boldsymbol{x}} \mathrm{d}\boldsymbol{k} \tag{5}$$

While the second one is defined as

$$\mathfrak{D}_{(2)}^{\chi} f = \int_{-\infty}^{+\infty} (ik)^{4\chi} \hat{f}(\boldsymbol{k}, t) e^{i\boldsymbol{k}\cdot\boldsymbol{x}} \mathrm{d}\boldsymbol{k} = \int_{-\infty}^{+\infty} e^{i2\pi\chi} k^{4\chi} \hat{f}(\boldsymbol{k}, t) e^{i\boldsymbol{k}\cdot\boldsymbol{x}} \mathrm{d}\boldsymbol{k} \tag{6}$$

To be compatible with ordinary diffusions of velocity and scalar and reserve a consistent form to previous investigations (e.g. Boutros and Gibbon [15]), the velocity and scalar diffusion terms are investigated by the second subset $\mathfrak{D}_{(2)}^{\chi}$. In contrast, to be consistent with the initial definitions by Zhao [18,19], the forcing term is denoted with the first subset $\mathfrak{D}_{(1)}^{\chi}$.

*2.2 Conservative laws and models*

In this investigation, the governing equations of velocity ($\hat{\boldsymbol{u}}$) and scalar fluctuations ($\widehat{s'}$) regarding anomalous diffusions in the turbulence driven by multiscale force can be generally expressed as

$$\left(\frac{\mathrm{d}}{\mathrm{d}t}+\hat{\boldsymbol{u}}\cdot\boldsymbol{\nabla}\right)\hat{\boldsymbol{u}}=-\frac{1}{\hat{\rho}}\nabla\hat{p}+c_u\mathfrak{D}_{(2)}^{\frac{\gamma}{4}}\hat{\boldsymbol{u}}+\boldsymbol{M}\mathfrak{D}_{(1)}^{\frac{\beta}{4}}\widehat{s'} \tag{7a}$$

$$\left(\frac{\mathrm{d}}{\mathrm{d}t}+\hat{\boldsymbol{u}}\cdot\boldsymbol{\nabla}\right)\widehat{s'}=-\boldsymbol{N}\cdot\hat{\boldsymbol{u}}+c_s\mathfrak{D}_{(2)}^{\frac{\alpha}{4}}\widehat{s'} \tag{7b}$$

$$\boldsymbol{\nabla}\cdot\hat{\boldsymbol{u}}=\boldsymbol{0} \tag{7c}$$

where $\hat{\rho}$ is the referenced fluid density, $\hat{\boldsymbol{u}}=\hat{\boldsymbol{u}}(\boldsymbol{x},t)$ denotes velocity vector, $\hat{p}=\hat{p}(\boldsymbol{x},t)$ is pressure, $\widehat{s'}=\widehat{s'}(\boldsymbol{x},t)$ is the fluctuation of the scalar $\hat{s}$. $\boldsymbol{M}\mathfrak{D}_{(1)}^{\beta/4}\widehat{s'}$ is the multiscale force related to the scalar field $\widehat{s'}$ [17,18], with $\boldsymbol{M}$ being the dimensional vector associated with the physical field. $\boldsymbol{N}=\nabla\langle\hat{s}\rangle$ is the vector related to the gradient of mean scalar $\langle\hat{s}\rangle$, to characterize the feature of stratified scalar background. In Eq. (7b), we have assumed $\Delta^{\alpha}\langle\hat{s}\rangle\ll\Delta^{\alpha}\widehat{s'}$, i.e. the scalar background is sufficiently smooth.

In this model, the influence of anomalous diffusions of both momentum and scalar has been taken into account, in the forms of $c_u\mathfrak{D}_{(2)}^{\gamma/4}\hat{\boldsymbol{u}}$ and $c_s\mathfrak{D}_{(2)}^{\alpha/4}\widehat{s'}$ respectively. $c_u$ and $c_s$ are the corresponding diffusivity coefficients with dimensions of $L^{\gamma}/T$ and $L^{\alpha}/T$ respectively. When $\gamma=\alpha=2$, the model returns to ordinary diffusion, with $c_u$ and $c_s$ equivalent to kinematic viscosity and diffusivity of scalar respectively. When $\gamma$ or $\alpha$ is over 2, the momentum or scalar experiences subdiffusion, while $\gamma$ or $\alpha$ is below 2, the momentum or scalar experiences superdiffusion.

In Fourier space, the governing equations can be [23-27]

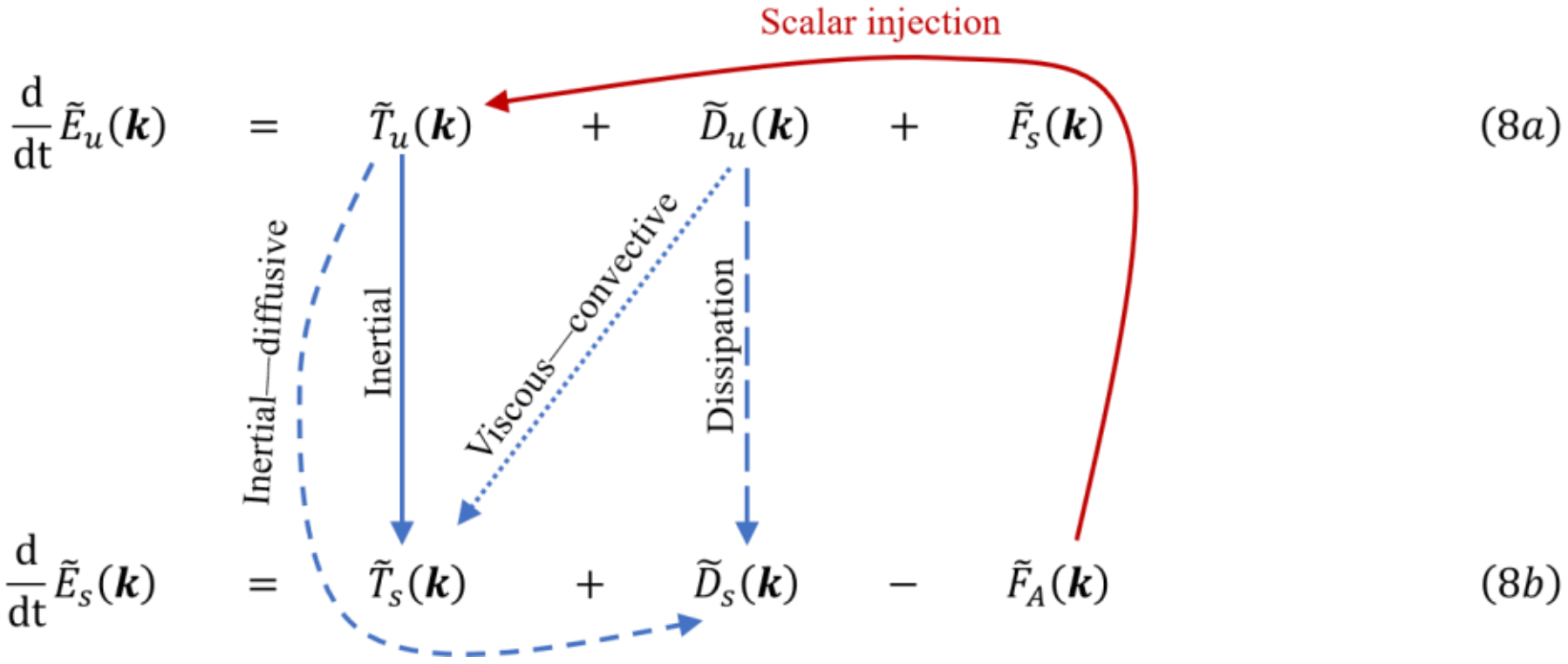

$$\frac{\mathrm{d}}{\mathrm{dt}}\tilde{E}_u(\boldsymbol{k}) = \tilde{T}_u(\boldsymbol{k}) + \widetilde{D}_u(\boldsymbol{k}) + \tilde{F}_s(\boldsymbol{k}) \tag{8a}$$

$$\frac{\mathrm{d}}{\mathrm{dt}}\tilde{E}_s(\boldsymbol{k}) = \tilde{T}_s(\boldsymbol{k}) + \widetilde{D}_s(\boldsymbol{k}) - \tilde{F}_A(\boldsymbol{k}) \tag{8b}$$

where $\tilde{E}_u(\boldsymbol{k})=|\boldsymbol{u}(\boldsymbol{k})|^2/2$ and $\tilde{E}_s(\boldsymbol{k})=|s'(\boldsymbol{k})|^2/2$ are the modal kinetic energy and scalar variance in wavenumber space, $\boldsymbol{u}$ and $s'$ are the Fourier transform of $\hat{\boldsymbol{u}}$ and $\widehat{s'}$ respectively. $\tilde{T}_u(\boldsymbol{k})$ and $\widetilde{D}_u(\boldsymbol{k})$ are the nonlinear kinetic energy transfer rate and dissipation rate corresponding to anomalous momentum diffusion, respectively. $\tilde{T}_s(\boldsymbol{k})$ and $\widetilde{D}_s(\boldsymbol{k})$ are the nonlinear transfer rate of the scalar variance and scalar dissipation rate corresponding to anomalous scalar diffusion, respectively. $\tilde{F}_s(\boldsymbol{k})$ denotes the energy feeding rate by the multiscale force due to the scalar field and $\tilde{F}_A(\boldsymbol{k})$ is the scalar feeding rate by bulk components. The lines with arrows show the possible transport routes in this flow. These quantities can be expressed as

$$\tilde{T}_u(\boldsymbol{k}) = \sum_{\boldsymbol{m}} \mathrm{Im}\{[\boldsymbol{k}\cdot\boldsymbol{u}(\boldsymbol{n})][\boldsymbol{u}(\boldsymbol{m})\cdot\boldsymbol{u}^*(\boldsymbol{k})]\} \quad (9a)$$

$$\tilde{T}_s(\boldsymbol{k}) = \sum_{\boldsymbol{m}} \mathrm{Im}\{[\boldsymbol{k}\cdot\boldsymbol{u}(\boldsymbol{n})][s'(\boldsymbol{m})s'^*(\boldsymbol{k})]\} \quad (9b)$$

$$\tilde{F}_s(\boldsymbol{k}) = k^\beta \mathrm{Re}[s'(\boldsymbol{k})\boldsymbol{M}\cdot\boldsymbol{u}^*(\boldsymbol{k})] \quad (9c)$$

$$\tilde{F}_A(\boldsymbol{k}) = \mathrm{Re}[s'(\boldsymbol{k})\boldsymbol{N}\cdot\boldsymbol{u}^*(\boldsymbol{k})] \quad (9d)$$

$$\tilde{D}_u(\boldsymbol{k}) = 2c_u k^\gamma \mathrm{Re}\left(e^{i\frac{1}{2}\pi\gamma}\right) E_u(\boldsymbol{k}) \quad (9e)$$

$$\tilde{D}_s(\boldsymbol{k}) = 2c_s k^\alpha \mathrm{Re}\left(e^{i\frac{1}{2}\pi\alpha}\right) E_s(\boldsymbol{k}) \quad (9f)$$

where Re and Im represent the real and imaginary parts of the quantity. $\boldsymbol{k} = \boldsymbol{m} + \boldsymbol{n}$. From Eqs. (9e, f), it can be seen the dissipation rate of kinetic energy and scalar variance are strictly determined by the orders of fractional derivations. Taking the dissipation of scalar variance as an example, both $k^\alpha$ and $\mathrm{Re}\left(e^{i\frac{1}{2}\pi\alpha}\right)$ control the magnitude and sign of $\tilde{D}_s$. For ordinary momentum diffusion where $\alpha = 2$, $\tilde{D}_s(\boldsymbol{k}) = -2c_s k^2 \tilde{E}_s(\boldsymbol{k})$ which is consistent to the negative value of conventional definitions. While for superdiffusion $\alpha = 3/2$, $\tilde{D}_s(\boldsymbol{k}) = -\sqrt{2}c_s k^{3/2}\tilde{E}_s(\boldsymbol{k})$ which shows a much slower variation with $k$ relative to the ordinary counterpart, implying a postponing influence towards higher $k$.

In a spherical shell around $k$ with a thickness of $dk$ in the wavenumber space, given the spectral kinetic energy flux $\Pi_u(k) = -\sum_{|\boldsymbol{k}'|\le k} \tilde{T}_u(\boldsymbol{k}')$ and the spectral scalar variance flux $\Pi_s(k) = -\sum_{|\boldsymbol{k}'|\le k} \tilde{T}_s(\boldsymbol{k}')$ [24,28], Eqs. (8a, b) under statistical equilibrium state become

$$\frac{\mathrm{d}}{\mathrm{d}k}\Pi_u(k) = F_s(k) + D_u(k) \quad (10a)$$

$$\frac{\mathrm{d}}{\mathrm{d}k}\Pi_s(k) = -F_A(k) + D_s(k) \quad (10b)$$

where $F_s(k)$ and $F_A(k)$ are the energy feeding rate by multiscale force due to the scalar field and the scalar feeding rate at wavenumber $k$, $D_u$ and $D_s$ are the corresponding dissipation terms respectively. They are

$$F_s(k) = k^\beta \sum_{|\boldsymbol{k}'|=k} \mathrm{Re}[s'(\boldsymbol{k}')\boldsymbol{M}\cdot\boldsymbol{u}^*(\boldsymbol{k}')] \quad (11a)$$

$$F_A(k) = \sum_{|\boldsymbol{k}'|=k} \mathrm{Re}[s'(\boldsymbol{k}')\boldsymbol{N}\cdot\boldsymbol{u}^*(\boldsymbol{k}')] \quad (11b)$$

$$D_u(k) = 2c_u k^\gamma \mathrm{Re}\left(e^{i\frac{1}{2}\pi\gamma}\right) E_u(k) \quad (11c)$$

$$D_s(k) = 2c_s k^\alpha \mathrm{Re}\left(e^{i\frac{1}{2}\pi\alpha}\right) E_s(k) \quad (11d)$$

where $E_u(k)$ is the 1D power spectrum of kinetic energy, $E_s(k)$ is the 1D power spectrum of scalar variance. For simplicity, we restrict ourselves to the case where the mean scalar gradient and the external force field are parallel, thus Eqs (11a, b) give

$$F_s(k) - \frac{M}{N}F_A(k)k^\beta = 0 \quad (12)$$

where $M = |\boldsymbol{M}|$ and $N = |\boldsymbol{N}|$ respectively. In a 3D turbulent flow, phenomenologically [28],

$$E_u(k) = u_k^2/k \sim k^{\xi_u} \quad (13a)$$

$$E_s(k) = s_k^2/k \sim k^{\xi_s} \quad (13b)$$

$$\Pi_u(k) = k u_k^3 \sim k^{\lambda_u} \quad (13c)$$

$$\Pi_s(k) = k s_k^2 u_k \sim k^{\lambda_s} \quad (13d)$$

where $\xi_u$, $\xi_s$, $\lambda_u$, $\lambda_s$ denote the scaling exponents of $E_u$, $E_s$, $\Pi_u$ and $\Pi_s$ respectively, $u_k$ and $s_k$ represent the spectral components of velocity and scalar. $u_k$ has a dimension of $\sim L/T$ and $s_k$ has a dimension of $\sim 1$. They are different from those of $\boldsymbol{u}$ and $s'$. Given $\Pi_u$ and $\Pi_s$ are the fluxes regarding nonlinear velocity and scalar transports, Eqs. (13c, d) are applicable only in the region where nonlinearity is dominant. Thus, they are not applicable or will cause serious deviation in the dissipation subranges of both velocity and scalar. From Eqs. (10a, b), it can be inferred that in the inertial subrange of 3D turbulence where the forcing and dissipation terms are negligibly small, thus $\mathrm{d}\Pi_u/\mathrm{d}k = 0$ and $\mathrm{d}\Pi_s/\mathrm{d}k = 0$. The inertial subrange must have constant $\Pi_u$ and $\Pi_s$. This gives $u_k \sim k^{-1/3}$ and $s_k \sim k^{-1/3}$.

In the subrange where the transports of kinetic energy and scalar variance are dominated by the multiscale force, we call it multiscale-force dominated (MFD) subrange [18]. In buoyancy-driven turbulence, it is the regime where BO59 law is established [29,30]. While in electrokinetic turbulence, it is corresponding to the EBF-dominant subrange [19,31]. According to the literatures [17,18], the structure of turbulence is strictly relied on $\beta$, with $\beta = 2/3$ being a critical value. When $\beta < 2/3$, the MFD subrange locates on the low-wavenumber side of the inertial subrange (Fig. 1). While $\beta > 2/3$, the MFD subrange locates on the high-wavenumber side of the inertial subrange and directly interacts with the dissipation subranges. In this investigation, we only discuss the cases without forcing and with external forcing of $\beta < 2/3$. The case $\beta > 2/3$ will be discussed in a sequential manuscript as Part 2.

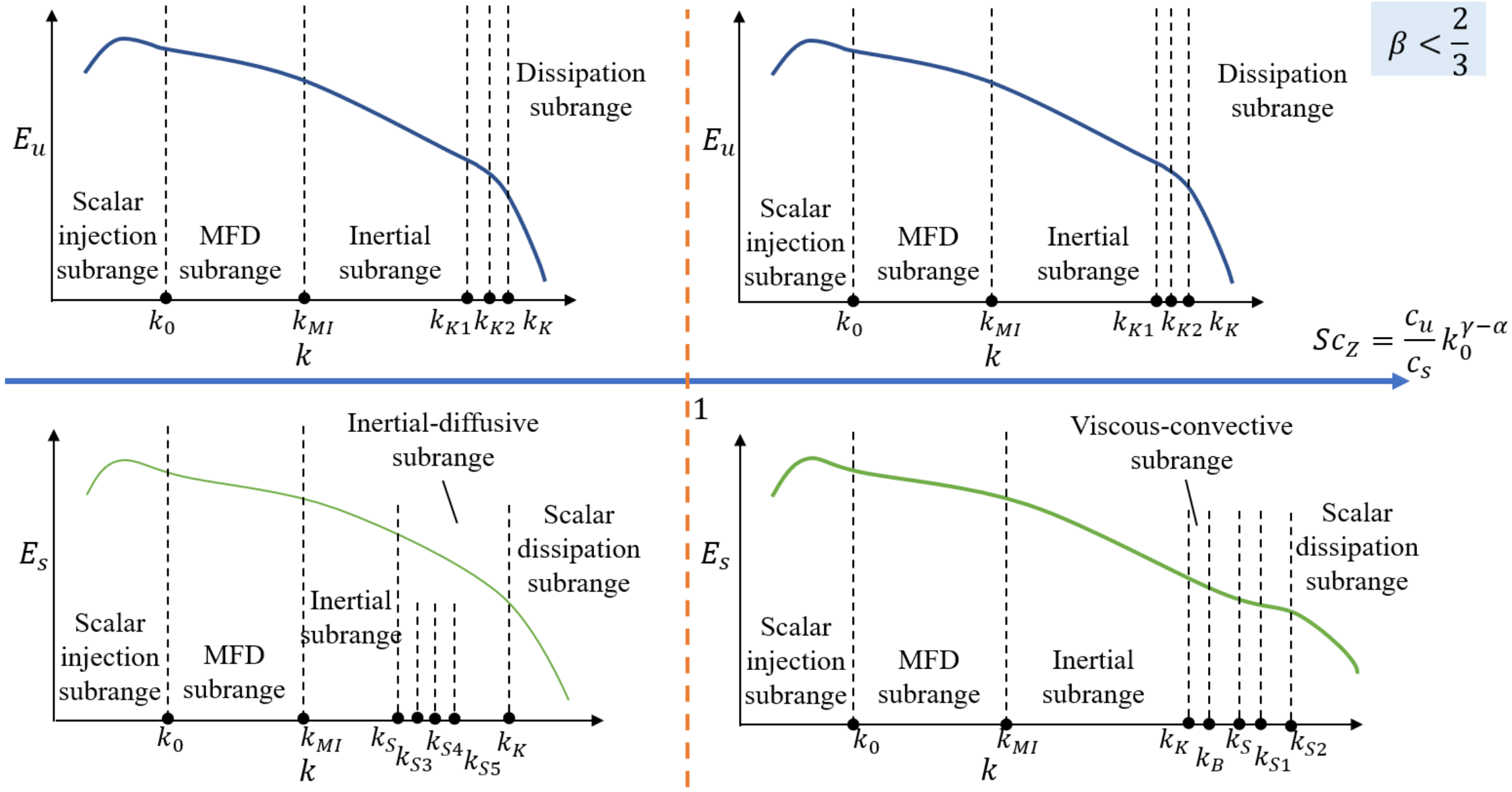

Fig. 1 Fine structures of momentum-scalar coupling turbulence in the no forcing or long-range forcing regime ($\beta < 2/3$). The wavenumbers $k_B$, $k_S$, $k_{S1}$ and $k_{S2}$ are plotted for the ordinary momentum diffusion ($\gamma = 2$) but anomalous scalar diffusion ($\alpha = 2.5$), while $k_0$, $k_K$ and $k_{MI}$ are shown for the general case.

When the turbulent kinetic energy is generated by the disturbing of volume force in the MFD subrange, it directly cascades along wavenumber into inertial subrange where $\Pi_u$ becomes constant. The MFD and inertial subrange are separated by a characteristic wavenumber $k_{MI}$. In buoyancy-driven turbulence, Bolgiano scale [32] is corresponding to one type of $k_{MI}$. As $k$ keeps increasing, dissipation of turbulent kinetic energy becomes dominant, forming dissipation subrange locates on the high-wavenumber side of the inertial subrange. From Eq. (11c), it can be inferred that, if $\gamma$ is decreased, the dissipation capability of fluid viscosity on turbulent kinetic energy can be significantly reduced. The weaker dissipation capability should result in a larger characteristic wavenumber $k_K$ at which the inertial subrange and dissipation subrange intersects.

Accompanied with the cascade of turbulent kinetic energy, the cascade of scalar variance also experiences three subranges, even if no scalar is produced during the process. The scalar variance is injected from large scales or mean field. It cascades along the MFD subrange, inertial subrange (constant $\Pi_s$), and then smeared in the scalar dissipation subrange, which is significantly influenced by $\alpha$. When $\alpha$ is reduced, the scalar dissipation can be weakened as well, resulting in a much larger characteristic wavenumber $k_S$ where scalar dissipation becomes dominant.

**3. Kinetic energy spectra and characteristic scales**

For the cases of momentum-scalar coupling turbulent with $\beta < 2/3$ or free decaying turbulence without forcing, the inertial subranges and dissipation subranges for turbulent kinetic energy and scalar variance intersect at $k_K$ and $k_S$ respectively. These two characteristic wavenumbers can be predicted phenomenologically from the convection and diffusion terms in Eqs. (7a) and (7b) separately, as

$$u_k(k_K)k_K^{1-\gamma} = c_u \tag{14a}$$

$$u_k(k_S)k_S^{1-\alpha} = c_s \tag{14b}$$

In inertial subrange where $E_u(k) = \varepsilon_u^{2/3}k^{-5/3}$, $E_s(k) = \varepsilon_u^{-1/3}\varepsilon_s k^{-5/3}$, it can be deduced from Eqs. (13a, b) that $u_k = \varepsilon_u^{1/3}k^{-1/3}$ and $s_k = \varepsilon_u^{-1/6}\varepsilon_s^{1/2}k^{-1/3}$, where $\varepsilon_u = |\int D_u \mathrm{d}k|$ and $\varepsilon_s = |\int D_s \mathrm{d}k|$ are the overall dissipations of kinetic energy and scalar variance. Thus, from Eqs. (14a, b), we have

$$k_K = \left(\frac{\varepsilon_u^{\frac{1}{3}}}{c_u}\right)^{\frac{1}{\gamma - \frac{2}{3}}} \tag{15a}$$

$$k_S = \left(\frac{\varepsilon_u^{\frac{1}{3}}}{c_s}\right)^{\frac{1}{\alpha - \frac{2}{3}}} \tag{15b}$$

It can be found in Eqs. (15a, b) that $\gamma = 2/3$ and $\alpha = 2/3$ are not applicable since the denominator vanishes. This observation is consistent to the investigation by Boutros and Gibbon [15] on fractional Navier-Stokes equation, where they found a critical exponent of 1/3 using fractional Laplacian operator. While the observation is mathematically interesting, this marginal case is of limited physical relevance for the anomalous diffusion, and thus, neglected in this investigation. If both diffusions are ordinary, i.e. $\gamma = \alpha = 2$, it can be calculated that $k_K = (\varepsilon_u/c_u^3)^{1/4}$ and $k_S = (\varepsilon_u/c_s^3)^{1/4}$. The former is the reciprocal of well-known Kolmogorov scale, while the latter is the reciprocal of scalar dissipation scale. However, if $\gamma$ and $\alpha$ both depart from 2, e.g. $\gamma = 5/3$ for superdiffusion, $k_K = (\varepsilon_u/c_u^3)^{1/3}$ which is definitely larger than that of ordinary diffusion. This is coincident to our prediction in last section. Therefore, it can be concluded that superdiffusion can enlarge $k_K$ and $k_S$, while subdiffusion reduces them.

Since the MFD subrange locates at low-wavenumber side of the inertial subrange, it is not direct interacted with dissipation. Therefore, the energy and scalar feeding terms ($F_s$ and $F_A$) which are related to the MFD subrange are negligible in the study on the influence of anomalous diffusions. Therefore, after substituting Eqs. (11c, d) into Eqs. (10a, b), we have

$$\frac{\mathrm{d}}{\mathrm{d}k}\Pi_u = 2c_u k^{\gamma}\mathrm{Re}\left(e^{i\frac{1}{2}\pi\gamma}\right)E_u \tag{16a}$$

$$\frac{\mathrm{d}}{\mathrm{d}k}\Pi_s = 2c_s k^{\alpha}\mathrm{Re}\left(e^{i\frac{1}{2}\pi\alpha}\right)E_s \tag{16b}$$

From Eqs. (13a-d), it can be inferred $\Pi_u \sim E_u^{\frac{3}{2}}k^{\frac{5}{2}} = C_{z_1}E_u^{\frac{3}{2}}k^{\frac{5}{2}}$ and $\Pi_s \sim E_s E_u^{\frac{1}{2}}k^{\frac{5}{2}} = C_{s_1}E_s E_u^{\frac{1}{2}}k^{\frac{5}{2}}$. $C_{z_1}$ and $C_{s_1}$ are dimensionless constants that are qualitatively related to the intermittency of velocity and scalar structures. They should be determined experimentally. Thus, after substituting $\Pi_u$ and $\Pi_s$ into Eqs. (16a, b), we have

$$\frac{\mathrm{d}}{\mathrm{d}k}E_u^{\frac{3}{2}}k^{\frac{5}{2}} = 2C_{z_1}^{-1}c_u k^{\gamma}\mathrm{Re}\left(e^{i\frac{1}{2}\pi\gamma}\right)E_u \tag{17a}$$

$$\frac{\mathrm{d}}{\mathrm{d}k}E_s E_u^{\frac{1}{2}}k^{\frac{5}{2}} = 2C_{s_1}^{-1}c_s k^{\alpha}\mathrm{Re}\left(e^{i\frac{1}{2}\pi\alpha}\right)E_s \tag{17b}$$

The turbulent kinetic energy spectrum equation Eq. (17a) is solved first. By the aid of Deepseek V3, we get a solution of $E_u$ as

$$\begin{aligned}E_u &= C_0^2 k^{-\frac{5}{3}} + \frac{4C_0 c_u}{(3\gamma-2)C_{z_1}}\mathrm{Re}\left(e^{i\frac{1}{2}\pi\gamma}\right)k^{\gamma-\frac{7}{3}} + \left[\frac{2c_u}{(3\gamma-2)C_{z_1}}\mathrm{Re}\left(e^{i\frac{1}{2}\pi\gamma}\right)\right]^2 k^{2\gamma-3} \\ &= C_K\varepsilon_u^{\frac{2}{3}}k^{-\frac{5}{3}}\left[1+\left(\frac{k}{k_{K1}}\right)^{\gamma-\frac{2}{3}}+\left(\frac{k}{k_{K2}}\right)^{2\gamma-\frac{4}{3}}\right], \qquad \gamma \neq \frac{2}{3}\end{aligned} \tag{18}$$

Here, Eq. (18) can be manually validated by substituting back into Eq. (17a). $C_0 = -C_K^{1/2}\,\varepsilon_u^{1/3}$ is determined to be consistent to the classic result, i.e. $E_u = C_K\varepsilon_u^{2/3}k^{-5/3}$ in the inertial subrange of turbulence at $\gamma = 2$. $C_K$ is Kolmogorov constant, $k_{K1}$ and $k_{K2}$ are two characteristic wavenumbers for the cases $\gamma \neq 2/3$ as

$$k_{K1} = \left[-\frac{C_{z_1}C_K^{\frac{1}{2}}(3\gamma-2)}{4\mathrm{Re}\left(e^{i\frac{1}{2}\pi\gamma}\right)}\right]^{\frac{1}{\gamma-\frac{2}{3}}}k_K \tag{19}$$

which is the intersection point of $k^{-5/3}$ and $k^{\gamma-7/3}$ subranges, and

$$k_{K2}=\left[-\frac{C_{z_1}C_K^{\frac{1}{2}}(3\gamma-2)}{2\mathrm{Re}\left(e^{i\frac{1}{2}\pi\gamma}\right)}\right]^{\frac{1}{\gamma-\frac{2}{3}}}k_K=2^{\frac{3}{3\gamma-2}}k_{K1} \tag{20}$$

which is the intersection point of $k^{\gamma-7/3}$ and $k^{2\gamma-3}$ subranges. For ordinary diffusion of momentum, i.e. $\gamma=2$, Eq. (18) becomes

$$E_u=\frac{c_u^2}{4C_{z_1}^2}k+\frac{C_K^{\frac{1}{2}}}{C_{z_1}}\varepsilon_u^{\frac{1}{3}}c_u k^{-\frac{1}{3}}+C_K\varepsilon_u^{\frac{2}{3}}k^{-\frac{5}{3}}=C_K\varepsilon_u^{\frac{2}{3}}k^{-\frac{5}{3}}\left[1+\left(\frac{k}{k_{K1}}\right)^{\frac{4}{3}}+\left(\frac{k}{k_{K2}}\right)^{\frac{8}{3}}\right] \tag{21}$$

In the current case, $k_{K1}=C_{z_1}^{3/4}C_K^{3/8}k_K$ and $k_{K2}=2^{3/4}k_{K1}$. If $C_{z_1}<1$, e.g. approximately 0.3 as shown later, both $k_{K1}$ and $k_{K2}$ can be smaller than $k_K$. Relative to the inertial subrange with $k^{-5/3}$, in the range $k\gg k_{K1}$, an elevation of $E_u$ can be predicted. This inspired us the possible explanation towards the well-known bottleneck effect. However, since Eq. (13c) is not applicable in the dissipation subrange, the flux deviation $\Pi_u-\varepsilon_u$ on the dissipation rate is not taken into account. Thus, the subsequent dissipation subrange of turbulent kinetic energy is not predicted in Eqs. (18) and (21).

To construct a comprehensive cascade route of turbulent kinetic energy towards dissipation subrange, the form of $E_u$ in the dissipation subrange regarding anomalous diffusion should be derived as well. Here, an approximation analysis is further conduct. We assume $D_u$ is a function of $\Pi_u$, say $D_u=\mathcal{F}(\Pi_u)$, it can be written after Taylor's expansion as

$$D_u(k)=\sum_{i=0}^{\infty}\frac{1}{i!}\frac{\mathrm{d}^i\mathcal{F}(\Pi_u)}{\mathrm{d}(\Pi_u)^i}\bigg|_{\Pi_{u,0}}\left(\Pi_u-\Pi_{u,0}\right)^i \tag{22}$$

where the initial value of spectral kinetic energy flux $\Pi_{u,0}=\varepsilon_u$ at $k=k_{MI}$, with $k_{MI}$ being the start point of inertial subrange (also the intersection wavenumber of MFD subrange and inertial subrange under forcing). Regarding a linear approximation from Eq. (22) and the more fundamental Eq. (11c), we have $D_u\sim\Pi_u$ and dimensionally $\Pi_u=C_{z_2}\varepsilon_u^{1/3}k^{5/3}E_u$ [33]. Thus, Eq. (16a) can be approximated as

$$\frac{\mathrm{d}}{\mathrm{d}k}\left(C_{z_2}\varepsilon_u^{\frac{1}{3}}k^{\frac{5}{3}}E_u\right)=2c_u k^{\gamma}\mathrm{Re}\left(e^{i\frac{1}{2}\pi\gamma}\right)E_u \tag{23}$$

By solving Eq. (23), we have $E_u=C_K\varepsilon_u^{\frac{2}{3}}k^{-\frac{5}{3}}\exp\left[2C_{z_2}^{-1}\mathrm{Re}\left(e^{i\frac{1}{2}\pi\gamma}\right)\left(\gamma-\frac{2}{3}\right)^{-1}(k/k_K)^{\gamma-\frac{2}{3}}\right]$. Although this solution cannot provide the correction term in Eq. (18), it is simple to infer the exponential form of $E_u$ in the dissipation range as we expect. When $\gamma=2$, it returns to Pao's theory [33] if $C_{z_2}^{-1}=C_K$. Finally, combine it and Eq. (18), an approximation solution of $E_u$ can be obtained empirically as

$$E_u=C_K\varepsilon_u^{\frac{2}{3}}k^{-\frac{5}{3}}\left[1+\left(\frac{k}{k_{K1}}\right)^{\gamma-\frac{2}{3}}+\left(\frac{k}{k_{K2}}\right)^{2\gamma-\frac{4}{3}}\right]\exp\left[\frac{2C_K\mathrm{Re}\left(e^{i\frac{1}{2}\pi\gamma}\right)}{\gamma-\frac{2}{3}}\left(\frac{k}{k_K}\right)^{\gamma-\frac{2}{3}}\right] \tag{24}$$

Remaining only the leading order of $\Pi_u$ yields a closed equation for $\Pi_u$ that incorporates a correction to the constant-flux state. This gives the general solution in Eq. (24) regarding dissipation. The exponential term in Eq. (24) thus represents the cumulative effect of the flux deviation on the spectral shape in the dissipation range. From Eq. (24), the influence of $\gamma$ on the cascade of turbulent kinetic energy is clear.

First of all, Fig. 2(a) shows how $k_K$ and $k_S$ vary with $\gamma$ and $\alpha$ respectively. Under the given $\varepsilon_u$, $k_K$ significantly decreases with $\gamma$ from 4/3 to 8/3, spanning up to $10^8$. Although $\gamma < 2$ is corresponding to superdiffusion, it shows a

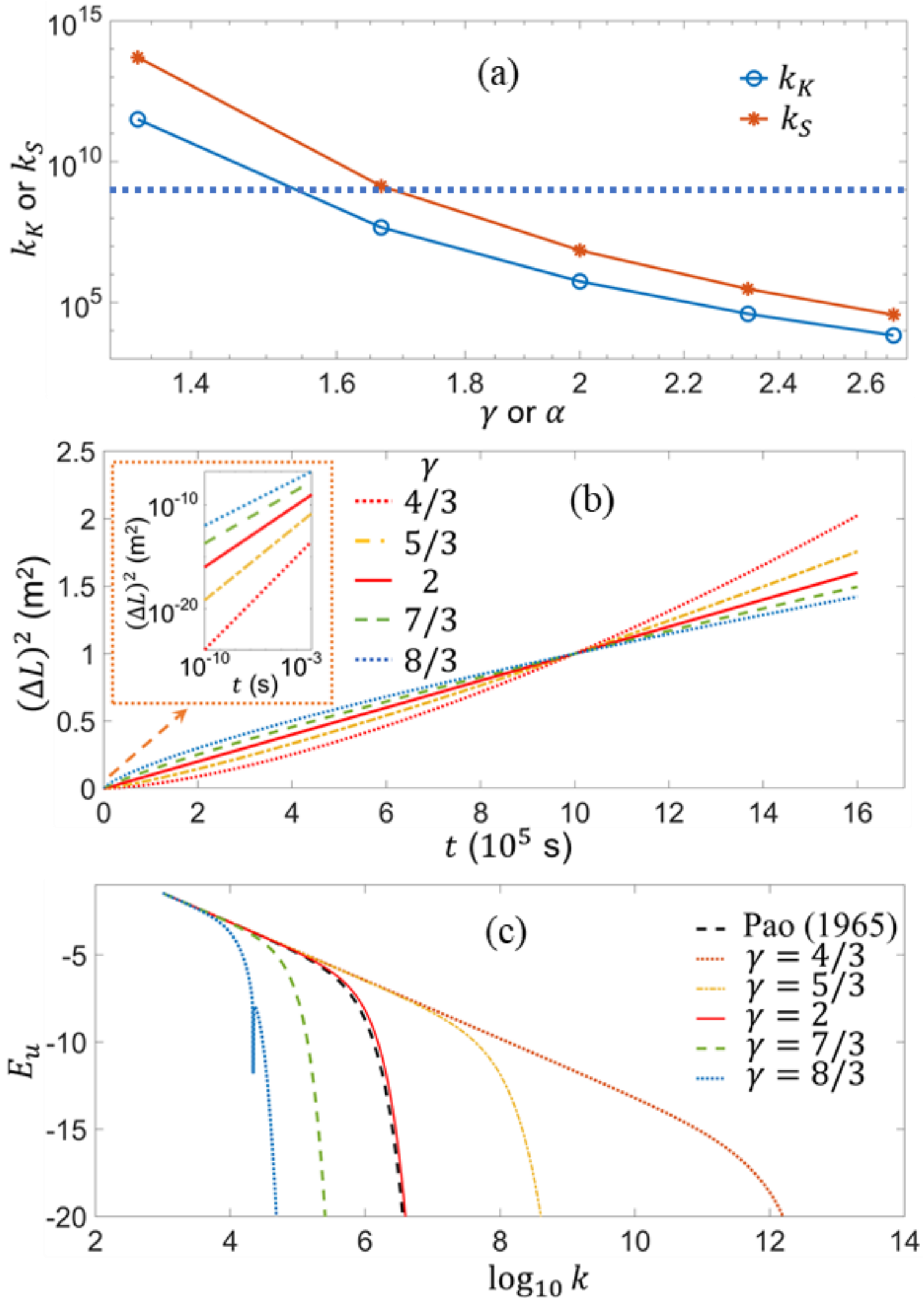

Fig. 2 (a) Characteristic wavenumbers $k_K$ (solid lines) and $k_S$ (dashed lines) as functions of $\gamma$ and $\alpha$, respectively; (b) mean-square displacement $(\Delta L)^2$ versus time $t$ for different $\gamma$, illustrating the crossover from superdiffusive ($\gamma < 2$) to ordinary diffusive ($\gamma = 2$) behaviour; inset: early-time behaviour showing slower transport for superdiffusion at small $t$; (c) schematic of the inertial-range broadening effect: decreasing $\gamma$ shifts $k_K$ to higher wavenumbers, widening the inertial subrange. Here, $\varepsilon_u = 10^5$ m$^2$/s$^3$, $\varepsilon_s = 10^4$ 1/s$^3$, $c_u = 10^{-6}$ m$^\gamma$/s, $c_s = 5 \times 10^{-10}$ m$^\alpha$/s.

faster transport of momentum at longer time, as can be seem from the mean-square displacement $(\Delta L)^2$ in Fig. 2(b). It is much slower in momentum transport at small time ($t < (\Delta L_c)^\gamma / c_u$ with $\Delta L_c = 1$ m) and length scales (inset of Fig. 2(b)), relative to ordinary diffusion counterpart. This is counter-intuitive at first glance. It leads to a much wider inertial subrange, as can be found in Fig. 2(c). The turbulent kinetic energy can be delivered into an unpredicted high wavenumber. Thus, a highly turbulent state is more reachable at a smaller $\gamma$.

When $\gamma = 2$, Eq. (24) has the following form

$$E_u = \left( C_K \varepsilon_u^{\frac{2}{3}} k^{-\frac{5}{3}} + \frac{C_K^{\frac{1}{2}}}{C_{z_1}} \varepsilon_u^{\frac{1}{3}} c_u k^{-\frac{1}{3}} + \frac{c_u^2}{4C_{z_1}^2} k \right) \exp\left[ -\frac{3}{2} C_K \left( \frac{k}{k_K} \right)^{\frac{4}{3}} \right]$$
$$= C_K \varepsilon_u^{\frac{2}{3}} k^{-\frac{5}{3}} \left[ 1 + \left( \frac{k}{k_{K1}} \right)^{\frac{4}{3}} + \left( \frac{k}{k_{K2}} \right)^{\frac{8}{3}} \right] \exp\left[ -\frac{3}{2} C_K \left( \frac{k}{k_K} \right)^{\frac{4}{3}} \right] \tag{25}$$

Relative to the classical Pao's theory [33] where $E_u = C_K \varepsilon_u^{2/3} k^{-5/3} \exp\left[ -\frac{3}{2} C_K \left( \frac{k}{k_K} \right)^{4/3} \right]$, Eq. (25) shows more fine structures as plotted with compensated kinetic energy spectrum in Fig. 3. Particularly, the term $k$ and term $k^{-1/3}$ cause deviation from Pao's theory, tending to inhibit the descending of $E_u$. With properly selecting $C_{z_1}$ around 0.3, e.g. 0.297 and 0.323 respectively, both the spectrum bumps and the bump locations are consistent to the numerical [34] and experimental [35] investigations on the bottleneck effect [36,37] in compensated kinetic energy spectrum. The former shows bump locates at $k/k_K = 0.153$ and the latter shows bump locates at $k/k_K \approx 0.04$~$0.05$, which are both reproduced in this model. Lots of investigations have shown the influence of viscosity and hyperviscosity [38-40] on the bottleneck effect. In fact, both the linear approximation of $D_u$ in Eq. (23) and changing $\gamma$ in Eq. (24) are equivalent to alternating the influence of viscosity and hyperviscosity, causing elevation of $E_u$ in high wavenumber region from the large $k^{\gamma-7/3}$ and $k^{2\gamma-3}$ terms. Relative to the influence of hyperviscosity, this model shows that the leading order influence of $\Pi_u$ on $D_u$ is more dominant on the presence of bottleneck effect, which is observable even without hyperviscosity. After the bottleneck region, the exponential term becomes dominant and $E_u$ decreases exponentially as the dissipation subrange.

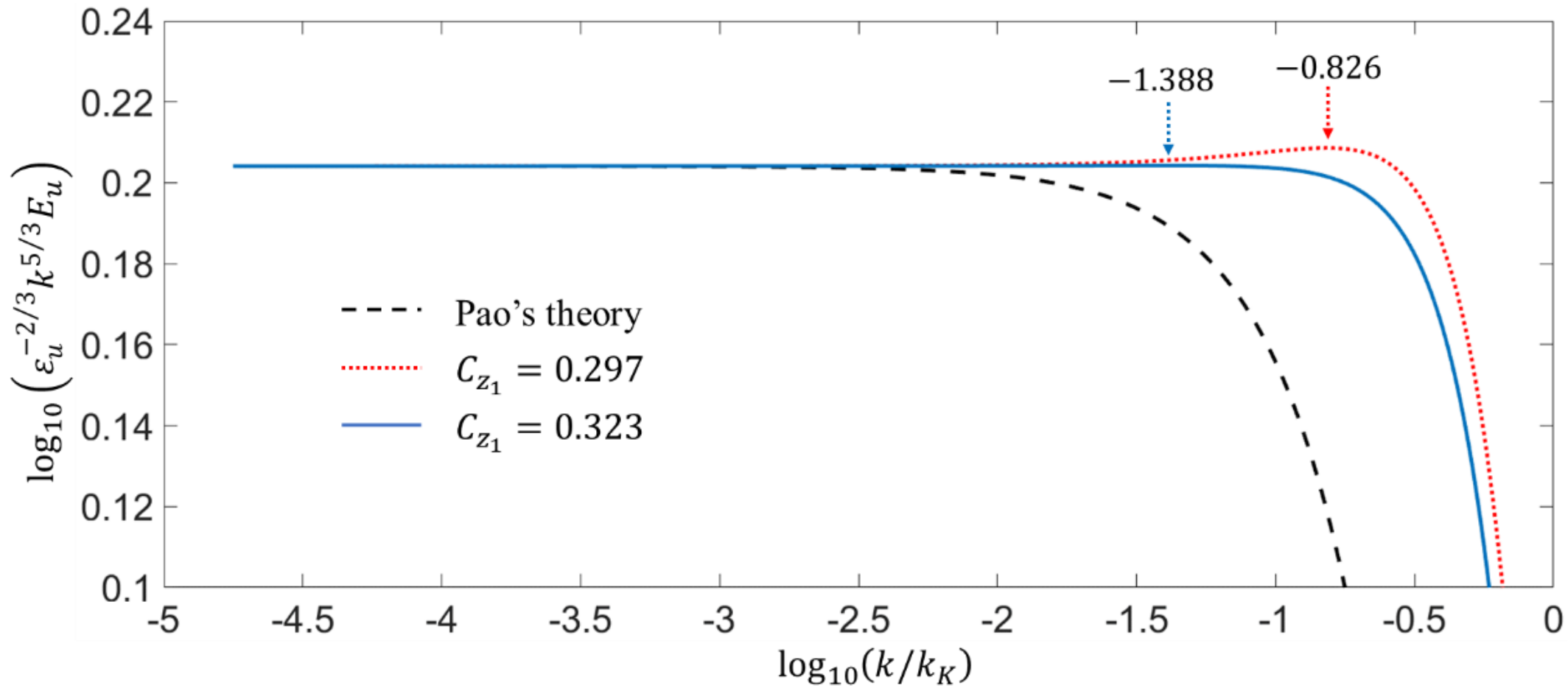

Fig. 3 Comparison of compensated kinetic energy spectra between Eq. (25) with Pao's theory for ordinary momentum diffusion. The blue dashed arrow ($k/k_K = 0.040$) and red dashed arrow ($k/k_K = 0.153$) point out the bump locations where the compensated kinetic energy spectra reach maximum. At $C_{z_1} = 0.297$, $k_{K1}/k_K = 0.480$ and $k_{K2}/k_K = 0.807$. At $C_{z_2} = 0.323$, $k_{K1}/k_K = 0.512$ and $k_{K2}/k_K = 0.861$.

## 4. Scalar spectra and characteristic scales at $Sc_Z \gg 1$

Relative to the spectral structure of turbulent kinetic energy, the scalar spectra have more and complex structures at high wavenumbers, depending on Schmidt number. Unlike the case in ordinary diffusions of momentum and scalar, where a Schmidt number can be simply defined as $Sc = c_u/c_s$, we cannot simply define a $Sc$ merely depending on the relationship among $c_u$, $c_s$, $\gamma$ and $\alpha$, since the dimensions cannot be balanced. In this investigation, an anomalous Schmidt number is defined as $Sc_Z = k_0^{\gamma-\alpha}\, c_u/c_s$, which describes the ratio of diffusion times of scalar and momentum on the minimum wavenumber (or the largest length scale). When $\gamma = \alpha$, it returns to the ordinary Schmidt number. When $\gamma > \alpha$, the momentum diffusion is more "subdiffusion" than the scalar diffusion. If $k_0 \ll 1$, the momentum diffusion is slow down by the relatively larger $\gamma$, in contrast to the scalar diffusion, resulting in a smaller $Sc_Z$. However, if $k_0 \gg 1$, it is easier to get $Sc_Z \gg 1$, i.e. a faster momentum diffusion than scalar diffusion. When $\gamma < \alpha$, the momentum diffusion is more "superdiffusion" than the scalar diffusion. If $k_0 \ll 1$, the momentum diffusion is promoted, in contrast to the scalar diffusion, resulting in a larger $Sc_Z$. On the other hand, if $k_0 \gg 1$, it is possible that $Sc_Z \ll 1$, i.e. a faster scalar diffusion than momentum diffusion. The relationship between $Sc_Z$ and $\gamma - \alpha$ can be found in Fig. 4.

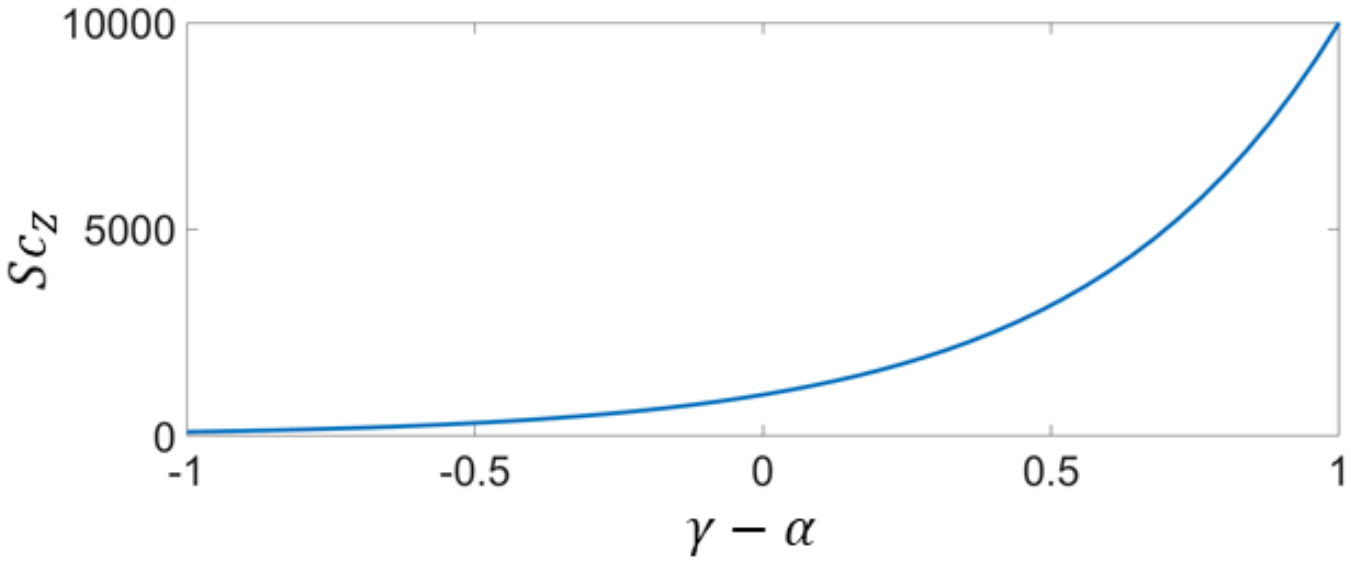

Fig. 4 Influence of $\gamma$, $\alpha$ on $Sc_Z$, where $k_0 = 10$ m$^{-1}$ and $c_u/c_s = 10^3$ m$^{\gamma-\alpha}$.

When $Sc_Z \gg 1$, the momentum diffusion is much faster than the scalar diffusion. The scalar structures can cascade into much smaller length scales (or larger wavenumber), leading to additional scaling subranges in the range of $k_K \ll k \ll k_S$.

First, in the inertial subrange of velocity where $k_{MI} \ll k \ll k_{K1}$, $E_u = C_K \varepsilon_u^{2/3} k^{-5/3}$. $E_s$ can be solved theoretically from Eq. (17b) as

$$E_s = C_p \varepsilon_u^{-\frac{1}{3}} \varepsilon_s k^{-\frac{5}{3}} \exp\left[\frac{2\mathrm{Re}\left(e^{i\frac{1}{2}\pi\alpha}\right)}{\left(\alpha - \frac{2}{3}\right)\sqrt{C_K}} \left(\frac{k}{k_S}\right)^{\alpha-\frac{2}{3}}\right], \quad \alpha \neq \frac{2}{3} \tag{26}$$

where we can rewrite $k_S$ with $Sc_Z$, as $k_S = Sc_Z^{1/\left(\alpha-\frac{2}{3}\right)} k_K^{\left(\gamma-\frac{2}{3}\right)/\left(\alpha-\frac{2}{3}\right)} k_0^{(\alpha-\gamma)/\left(\alpha-\frac{2}{3}\right)} = Sc_Z^{1/\left(\alpha-\frac{2}{3}\right)} \left(\frac{k_0}{k_K}\right)^{(\alpha-\gamma)/\left(\alpha-\frac{2}{3}\right)} k_K$. For ordinary diffusions of momentum and scalars, i.e. $\gamma = 2$ and $\alpha = 2$, we have

$$E_s = C_p \varepsilon_u^{-\frac{1}{3}} \varepsilon_s k^{-\frac{5}{3}} \exp\left[-\frac{3}{2} C_K^{-\frac{1}{2}} \left(\frac{k}{k_S}\right)^{\frac{4}{3}}\right] \tag{27}$$

with $k_S = Sc_Z^{3/4} k_K$. Thus, the scalar spectrum returns to the classical Pao's model [33,41]. For the cases that $\gamma = \alpha > 2/3$, $k_S = Sc_Z^{1/\left(\gamma-\frac{2}{3}\right)} k_K$ is much larger than $k_K$, due to $Sc_Z \gg 1$. Thus, the exponential term is dominant only at a very large wavenumber $k \gg k_K$. This is far beyond the constraint in this section that $k$ is in the range $k_{MI} \ll k \ll k_{K1}$, where we can only observe $E_s = C_p \varepsilon_u^{-1/3} \varepsilon_s k^{-5/3}$ in the inertial subrange.

Second, if considering the second term in Eq. (17b) in the range of $k_{K1} \ll k \ll k_{K2}$, i.e. the subrange where $E_u = -\frac{4C_K^{1/2}\varepsilon_u^{1/3} c_u}{(3\gamma-2)C_{z_1}} \mathrm{Re}\left(e^{i\frac{1}{2}\pi\gamma}\right) k^{\gamma-7/3}$, $E_s$ is solved theoretically as

$$E_s = \begin{cases} C_{p2} \varepsilon_u^{-\frac{2+3\gamma}{12}} \varepsilon_s^{\frac{2+3\gamma}{4}} k^{-\frac{8+3\gamma}{6}} \exp\left[\dfrac{Q_{p2}}{\left(\alpha - \dfrac{2+3\gamma}{6}\right)} \left(\dfrac{k}{k_{S1}}\right)^{\alpha-\frac{2+3\gamma}{6}}\right], & \alpha \neq \dfrac{2+3\gamma}{6} \\ C_{p3} k^{\left[Q_{p_2}(k_0/k_K)^{(3\gamma-2)/6} Sc_Z^{-1}\right] - \frac{8+3\gamma}{6}}, & \alpha = \dfrac{2+3\gamma}{6} \end{cases} \tag{28}$$

where $k_{S1} = \left[Sc_Z^{-1} k_0^{\gamma-\alpha} k_K^{-\left(\frac{\gamma}{2}-\frac{1}{3}\right)}\right]^{-1/\left(\alpha-\frac{2+3\gamma}{6}\right)} = k_S^{(6\alpha-4)/(6\alpha-2-3\gamma)} k_K^{(2-3\gamma)/(6\alpha-2-3\gamma)}$ is the scalar characteristic wavenumber, $Q_{p2} = 2\mathrm{Re}\left(e^{i\frac{1}{2}\pi\alpha}\right) \Big/ \sqrt{-4(3\gamma-2)^{-1} C_{z_1}^{-1} C_K^{1/2} \mathrm{Re}\left(e^{i\frac{1}{2}\pi\gamma}\right)}$. Regarding $Sc_Z \gg 1$ and $k_0/k_K \ll 1$, when $\gamma > 2/3$, $Q_{p2}(k_0/k_K)^{(3\gamma-2)/6} Sc_Z^{-1} \ll 1$. Thus, $E_s \approx C_{p3} k^{-(8+3\gamma)/6}$ for $\alpha = (2+3\gamma)/6$, with $C_{p3} \approx C_{p2} \varepsilon_u^{-(2+3\gamma)/12} \varepsilon_s^{(2+3\gamma)/4}$. Since $k_{S1} \gg k_{K2}$, in the range $k_{K1} < k < k_{K2}$, we can only observe $E_s \sim k^{-(8+3\gamma)/6}$.

For ordinary diffusions of momentum and scalars, i.e. $\gamma = 2$ and $\alpha = 2$, we have

$$E_s = C_{p2} \varepsilon_u^{-\frac{2}{3}} \varepsilon_s^2 k^{-\frac{7}{3}} \exp\left[\frac{3}{2} Q_{p_2} \left(\frac{k}{k_{S1}}\right)^{\frac{2}{3}}\right] \tag{29}$$

with $k_{S1} = Sc_Z^{3/2} k_K = k_S^2/k_K$ and $Q_{p_2} = -2\sqrt{C_{z_1}/C_K^{1/2}}$. It can be seen, as $Sc_Z \gg 1$, $k_{S1} \gg k_S \gg k_{K2} > k_{K1}$. Theoretically, $k^{-7/3}$ can be extended to a very high wavenumber. However, on one hand, $k^{-7/3}$ spectrum in Eq. (29) could be covered up by the exponential term in Eq. (27), if $\varepsilon_s$ is not sufficiently large. On the other hand, $k_{K2} = 2^{3/4} k_{K1}$ indicates the subrange has a very limit bandwidth. Therefore, its existence in the subrange $k_{K1} < k < k_{K2}$ remains debating.

Third, in the subrange $k_{K2} \ll k \ll k_K$ where $E_u = \left[\frac{2c_u}{(3\gamma-2)C_{z_1}}\right]^2 \mathrm{Re}(e^{i\pi\gamma}) k^{2\gamma-3}$, $E_s$ can be solved theoretically from Eq. (17b) as

$$E_s = \begin{cases} C_{p4}\varepsilon_u^{-\frac{\gamma}{2}}\varepsilon_s^{\frac{3\gamma}{2}}k^{-(\gamma+1)}\exp\left\{\frac{(3\gamma-2)C_{z_1}}{(\alpha-\gamma)}\mathrm{Re}\left[e^{i\frac{1}{2}\pi(\alpha-\gamma)}\right]\left(\frac{k}{k_{S2}}\right)^{\alpha-\gamma}\right\}, & \alpha \neq \gamma \\ C_{p5}\varepsilon_u^{\frac{1}{2}[C_{z_1}(3\gamma-2)Sc_Z^{-1}-\gamma]}\varepsilon_s^{-\frac{3}{2}[C_{z_1}(3\gamma-2)Sc_Z^{-1}-\gamma]}k^{C_{z_1}(3\gamma-2)Sc_Z^{-1}-(\gamma+1)}, & \alpha = \gamma \end{cases} \tag{30}$$

where $k_{S2} = k_0 Sc_Z^{1/(\alpha-\gamma)}$. Regarding $Sc_Z \gg 1$, $C_{z_1}(3\gamma - 2)Sc_Z^{-1} \ll 1$. Thus, $E_s \approx C_{p5}\varepsilon_u^{-\gamma/2}\varepsilon_s^{3\gamma/2}k^{-(\gamma+1)}$ if $\alpha = \gamma$. For ordinary diffusions of momentum and scalars,

$$E_s \approx C_{p5}\varepsilon_u^{-1}\varepsilon_s^{3}k^{-3} \tag{31}$$

Thus, the $k^{-3}$ spectrum might be observed in the subrange $k_{K2} \ll k \ll k_K$. Unfortunately, the difference between $k_{K2}$ and $k_K$ may be too small for the $k^{-3}$ spectrum to be identified.

Besides, for ordinary diffusions of momentum and scalars with $Sc_Z \gg 1$, there should be a viscous-convective subrange in the wavenumber range $k_K \ll k \ll k_B = k_K Sc_Z^{1/2}$, before the exponentially decaying subrange (or scalar dissipation subrange), according to Batchelor [42,43]. It is predicted $E_s \sim k^{-1}$ in the viscous-convective subrange. However, Eqs. (16) are not compatible for predicting this subrange. In this case, we use Pao's model [33,41] to derive the viscous-convective subrange regarding anomalous diffusion. In the viscous-convective subrange, $\Pi_s = E_s k/\tau$, with $\tau = \left(c_u^{2/\gamma}/\varepsilon_u\right)^{1/\left(3-\frac{2}{\gamma}\right)}$ is the reciprocal of strain rate. Therefore, Eq. (17b) becomes

$$\frac{\mathrm{d}}{\mathrm{d}k}(E_s k) = 2c_s\tau\mathrm{Re}\left(e^{i\frac{1}{2}\pi\alpha}\right)k^{\alpha}E_s \tag{32}$$

The solution is

$$E_s(k) = \varepsilon_u^{\frac{1}{2}}\varepsilon_s^{-\frac{3}{2}}k^{-1}\exp\left[2\mathrm{Re}\left(e^{i\frac{1}{2}\pi\alpha}\right)\alpha^{-1}\left(\frac{k}{k_B}\right)^{\alpha}\right] \tag{33}$$

where $k_B = Sc_Z^{1/\alpha}k_0(k_K/k_0)^{\gamma/\alpha} = k_S(k_K/k_S)^{2/3\alpha} = (k_S/k_K)^{(3\alpha-2)/3\alpha}k_K$ is a characteristic wavenumber at the joint of viscous-convective subrange and dissipation subrange. Given $Sc_Z \gg 1$, on one hand, it can be seen $k_S \gg k_B$. On the other hand, when $\alpha = \gamma > 2/3$, $k_B/k_K = Sc_Z^{1/\alpha} \gg 1$, indicating the viscous-convective subrange predicted by Batchelor [42] is beyond inertial subrange. Particularly, for ordinary diffusions of momentum and scalars, $k_B = k_K Sc_Z^{1/2}$ which is exactly the characteristic wavenumber corresponding to Batchelor's scale [42]. The corresponding spectrum is

$$E_s(k) = \varepsilon_u^{\frac{1}{2}}\varepsilon_s^{-\frac{3}{2}}k^{-1}\exp\left[-\left(\frac{k}{k_B}\right)^2\right] \tag{34}$$

which reproduces Batchelor's theory [42].

To elucidate the relative magnitudes of the fine scales in scalar structures, an example of the magnitudes of these characteristic scalar wavenumbers varies with $\alpha$ has been plotted in Fig. 5. Here, it is set $\gamma = 2$. When $\alpha$ is increased, all the characteristic scalar wavenumbers, i.e. $k_B$, $k_S$, $k_{S1}$ and $k_{S2}$, are decreased. In the left part with smaller $\alpha$, we have $k_{S2} \gg k_{S1} \gg k_S \gg k_B \gg k_K$. This is consistent to the prediction of $Sc_Z \gg 1$. While in the right part with larger $\alpha$, we have $k_{S2} \ll k_{S1} \ll k_S \ll k_B \ll k_K$, which is inconsistent to the prediction of $Sc_Z \gg 1$. This might be attributed to the rapid anomalous diffusion (due to high $\alpha$) of scalar on small scale structures, leading to smaller characteristic scalar wavenumbers. Since $k_S$, $k_{S1}$ and $k_{S2}$ vary so fast with $\alpha$, they could be simply exceeding the continuum limit of incompressible fluids, and approaching or even exceeding the wavenumber corresponding to Planck length [44] at small $\alpha$. Consequently, we restrict our physical analysis to the wavenumber range $10^{-3} \le k \le 2\pi \times 10^9$ m$^{-1}$, consistent with the continuum-fluid hypothesis, as illustrated in Fig. 5. All subsequent discussions are restricted to this physically meaningful parameter regime.

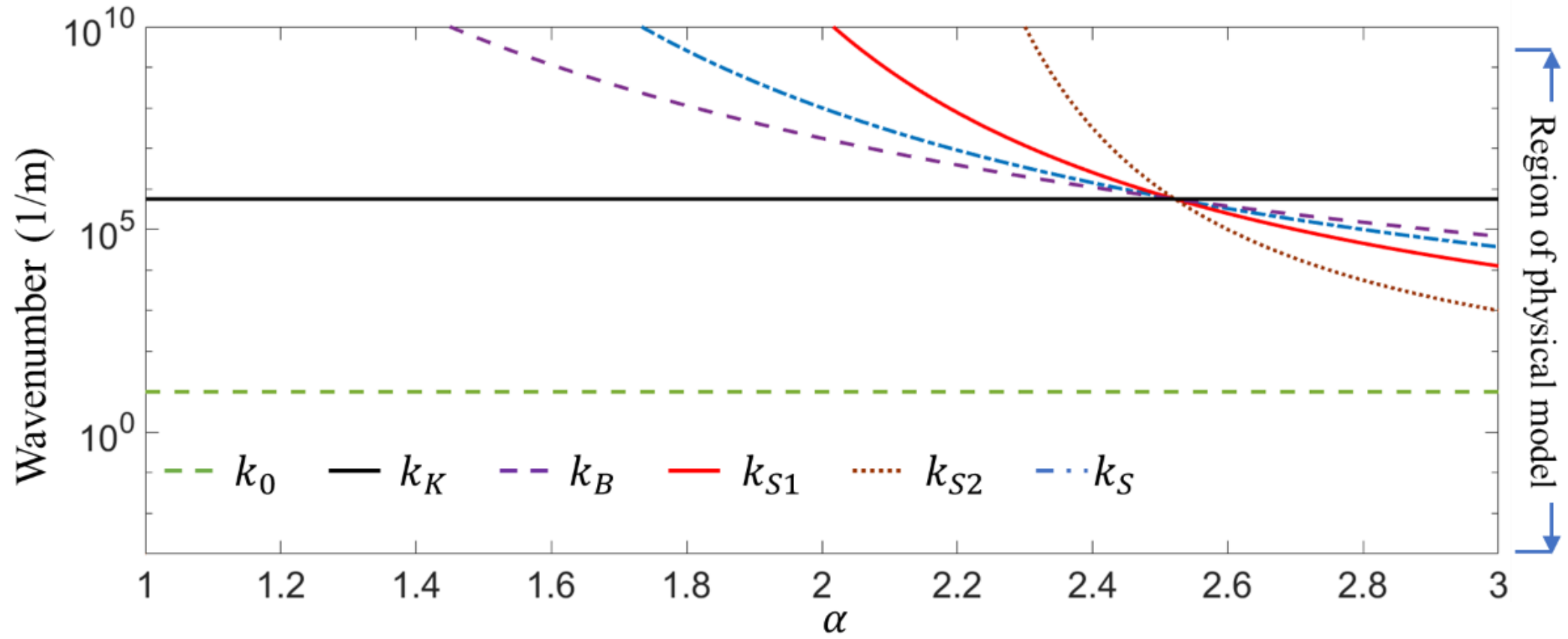

Fig. 5 Characteristic wavenumbers of scalar structures, where $k_0 = 10$ m$^{-1}$, $c_u = 10^{-6}$ m$^\gamma$, $c_s = 10^{-9}$ m$^\alpha$, $\varepsilon_u = 10^5$ m$^2$/s$^3$, and $\gamma = 2$.

### 5. Scalar spectra and characteristic scales at $Sc_Z \ll 1$

When $Sc_Z \ll 1$, the situation is more complex. There are two possible cases should be discussed separately.

(1) When $\gamma > \alpha$, for instance, $\gamma = 2$ and $\alpha = 3/2$, $k_B = Sc_Z^{2/3}(k_0/k_K)^{-1/3}k_K$ and $k_S = Sc_Z^{6/5}(k_0/k_K)^{-3/5}k_K$. In this case, even though $Sc_Z \ll 1$, it is still possible to have $k_S \gg k_K$ or $k_B \gg k_K$ according to different $k_0/k_K$, as can be seen from Fig. 6. This indicates the scalar structure can be transported into a larger wavenumber (or smaller scale) relative to velocity structure, even if $Sc_Z \ll 1$, showing significant difference from conventional turbulence which only considering ordinary diffusions of momentum and scalar.

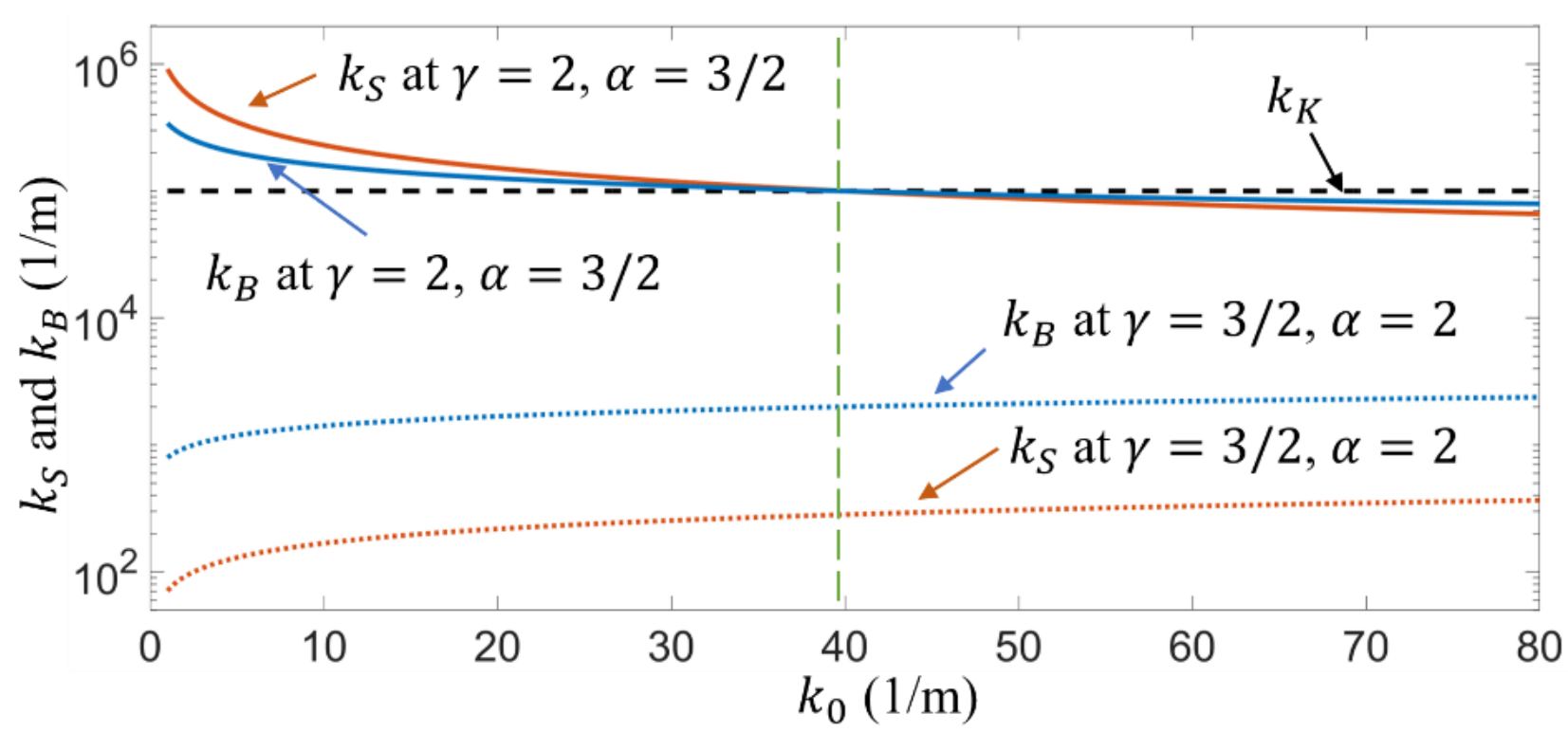

Fig. 6 Characteristic scalar wavenumbers $k_S$ and $k_B$ vary with $k_0$. Here, it is assumed $k_K = 10^5$ m$^{-1}$, $Sc_Z = 0.02$.

The reason can be attributed to that, $Sc_Z$ evaluate the relative diffusivity of momentum and scalar at the lowest wavenumber $k_0$. Although $Sc_Z \ll 1$ indicating the scalar diffusion is much stronger than momentum diffusion at $k_0$, it does not mean the scalar diffusion is much stronger than momentum diffusion at a larger wavenumber, e.g. $k_K$. For instance, if $c_u/c_s = 0.1$, $\gamma = 2$ and $\alpha = 3/2$, then $Sc_Z = k_0^{\gamma-\alpha}\, c_u/c_s = 0.32 < 1$ at $k_0 = 10$ m$^{-1}$. In contrast, at $k = 1000$ m$^{-1}$, $k^{\gamma-\alpha}\, c_u/c_s = 3.16 > 1$. This means, the momentum diffusion becomes faster than scalar diffusion on this scale. The scalar structure can be transported into a larger wavenumber relative to velocity structure even though $Sc_Z < 1$.

(2) When $\gamma \le \alpha$, for instance, $\gamma = 3/2$ and $\alpha = 2$, $k_B = Sc_Z^{1/2}(k_0/k_K)^{1/4}k_K$ and $k_S = Sc_Z^{3/4}(k_0/k_K)^{3/8}k_K$. In this case, since both $Sc_Z$ and $k_0/k_K$ are much smaller than unity, we have $k_S \ll k_B \ll k_K$, as can be seen in Fig. 6. Thus, there could exist an inertial-diffusive subrange according to Batchelor, Howells and Townsend [45], which should have a -17/3 scalar spectrum for ordinary diffusions. It, in fact, shows a triple interaction among the nonlinear velocity transport term ($\tilde{T}_u$), the nonlinear scalar transport term ($\tilde{T}_s$) and the scalar dissipation term ($\widetilde{D}_s$) in Eqs. (8a) and (8b). Some numerical simulations [46-48] have claimed that the spectrum could be observed at low Schmidt number. Later, Jolly and Wirosoetisno [49,50] provides strict mathematical analysis in both 2D and 3D to support the key hypothesis in Batchelor, Howells and Townsend [45]. In this part, we employ the phenomenological treatment by Batchelor, Howells and Townsend [45], i.e.

$$c_s^2 k^{2\alpha} E_s \sim E_u G_s \tag{35}$$

where $G_s = \langle |\nabla \widehat{s'}|^2 \rangle$ is the mean module of scalar gradient. Unlike in the case of ordinary diffusion, $G_s = \int k^2\, E_s dk$ is dimensionally different to $\varepsilon_s c_s^{-1} = \int k^\alpha \left|\mathrm{Re}\left(e^{i\frac{1}{2}\pi\alpha}\right)\right| E_s \mathrm{d}k$ in the case $\alpha \neq 2$. Thus, we don't have a simple and phenomenological expression on $G_s$ by giving $\varepsilon_s$, $c_s$ and $\alpha$. Given $G_s$ is a measurable quantity, after employing Eq. (18), dimensionally we have

$$\begin{aligned} E_s = C_K G_s c_s^{-2} \varepsilon_u^{2/3} k^{-\frac{5}{3}-2\alpha}\left[1+\left(\frac{k}{k_{K1}}\right)^{\gamma-\frac{2}{3}}+\left(\frac{k}{k_{K2}}\right)^{2\gamma-\frac{4}{3}}\right] &= C_K k_S^{-1}\left(\frac{k}{k_{S3}}\right)^{-\frac{5}{3}-2\alpha}\left[1+\left(\frac{k}{k_{K1}}\right)^{\gamma-\frac{2}{3}}+\left(\frac{k}{k_{K2}}\right)^{2\gamma-\frac{4}{3}}\right] \\ &= C_K k_S^{-1}\left[\left(\frac{k}{k_{S3}}\right)^{-\frac{5}{3}-2\alpha}+\left(\frac{k}{k_{S4}}\right)^{\gamma-2\alpha-\frac{7}{3}}+\left(\frac{k}{k_{S5}}\right)^{2\gamma-2\alpha-3}\right] \end{aligned} \tag{36}$$

where $k_{S3} = \left(G_s c_s^{-2-\frac{3}{3\alpha-2}} \varepsilon_u^{\frac{2}{3}+\frac{1}{3\alpha-2}}\right)^{1/\left(\frac{5}{3}+2\alpha\right)}$, $k_{S4} = \left(k_{S3}^{-\frac{5}{3}-2\alpha} k_{K1}^{\gamma-\frac{2}{3}}\right)^{1/\left(\gamma-2\alpha-\frac{7}{3}\right)}$, and $k_{S5} = \left(k_{S3}^{-\frac{5}{3}-2\alpha} k_{K2}^{2\gamma-\frac{4}{3}}\right)^{1/(2\gamma-2\alpha-3)}$ are the wavenumbers that characterize the scalar structure in inertial-diffusive subrange. The three terms in Eq. (36) are sequentially separated by all the characteristic wavenumbers.

In the case $Sc_Z \ll 1$, we plot these characteristic scalar wavenumbers in Fig. 7 and compare them with these characteristic wavenumbers of turbulent kinetic energy. Here, we make $k_K$, $k_{K1}$ and $k_{K2}$ unchanged. When $\alpha$ is increased, all the characteristic scalar wavenumbers (e.g. $k_S$, $k_{S3}$, $k_{S4}$ and $k_{S5}$) decrease rapidly (Fig. 7(a)). A critical point of $\alpha$ around 1.3 is observed. This critical point is not universal, but strictly relies on all the control parameters. In the region that physical model established, when $\alpha > 1.3$, we have $k_S \gg k_{S3} \gg k_{S4} \gg k_{S5}$. Except part of $k_S$, all the $k_{S3}$, $k_{S4}$ and $k_{S5}$ are smaller than $k_K$, $k_{K1}$ and $k_{K2}$. This means there are fine scalar structures according to Eq. (36) predicted in the inertial-diffusive subrange. However, when $\alpha < 1.3$, we have $k_{S5} \gg k_{S4} \gg k_{S3}$ which are all above $k_K$, $k_{K1}$ and $k_{K2}$. Therefore, in the inertial-diffusive subrange where $k \ll k_{K1}$, only the $k^{-\frac{5}{3}-2\alpha}$ spectrum can be retained.

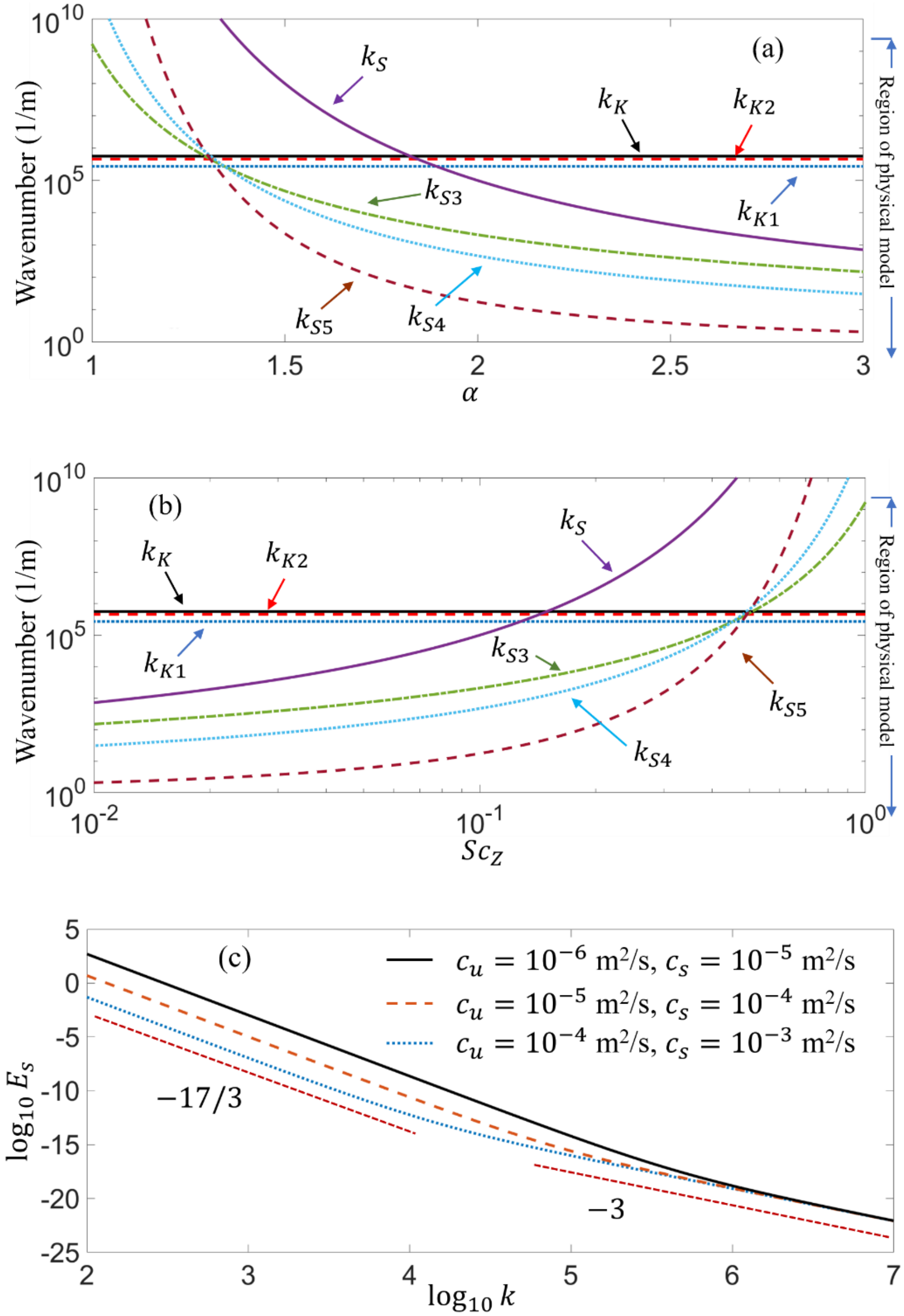

Fig. 7 Relationship among $k_{S3}$, $k_{S4}$ and $k_{S5}$ when $Sc_Z \ll 1$. Here, $C_{Z1} = 0.3$, $G_S = 3$ which is calculated according to [48] where the Schmidt number is 1/128, $k_0 = 10$ m$^{-1}$, $\varepsilon_u = 10^5$ m$^2$/s$^3$, and $\varepsilon_s = 10^2$ m$^2$/s$^3$. (a) Influence of $\alpha$, where $\gamma = 2$, $c_u = 10^{-6}$ m$^\gamma$, $c_s = 10^{-5}$ m$^\alpha$. (b) Influence of $Sc_Z$, where $\gamma = 2$, $c_u = 10^{-6}$ m$^\gamma$, $c_s = 10^{-5}$ m$^\alpha$. (c) $E_s$ in the inertial-diffusive subrange, where $\gamma = \alpha = 2$.

If $\gamma$ and $k_0$ are given, $Sc_Z$ is uniquely determined by $\alpha$. In Fig. 7(b), there also exists a critical point of $Sc_Z = 0.5$ correspondingly. In contrast to $\alpha$, $k_{S3}$, $k_{S4}$ and $k_{S5}$ increase with $Sc_Z$ rapidly, indicating the influence of reduced scalar diffusion. When $Sc_Z < 0.5$, $k_{S3} \gg k_{S4} \gg k_{S5}$ and they are all below $k_K$, $k_{K1}$ and $k_{K2}$. $E_s$ follows Eq. (36) in the inertial-diffusive subrange. In contrast, when $Sc_Z > 0.5$, $k_{S5} \gg k_{S4} \gg k_{S3}$ which are all above $k_K$, $k_{K1}$ and $k_{K2}$. Only the $k^{-\frac{5}{3}-2\alpha}$ spectrum can be predicted.

For ordinary diffusions of momentum and scalars, we get

$$E_s = C_K k_S^{-1} \left[ \left( \frac{k}{k_{S3}} \right)^{-\frac{17}{3}} + \left( \frac{k}{k_{S4}} \right)^{-\frac{13}{3}} + \left( \frac{k}{k_{S5}} \right)^{-3} \right] \tag{37}$$

where $k_{S3} = \left( G_s c_s^{-\frac{11}{4}} \varepsilon_u^{\frac{11}{12}} \right)^{3/17}$, $k_{S4} = \left( k_{S3}^{-\frac{17}{3}} k_{K1}^{\frac{4}{3}} \right)^{-3/13}$, and $k_{S5} = \left( k_{S3}^{-\frac{17}{3}} k_{K2}^{\frac{8}{3}} \right)^{-1/3}$. A schematic of $E_s$ has been plotted in Fig. 7(c). Several observations can be concluded. First, the -17/3 scalar spectrum by Batchelor, Howells and Townsend [45] was recovered in the inertial-diffusive subrange at low wavenumber region. Second, a -3 scalar spectrum was found at the low wavenumber region, if no scalar dissipation is regarded. Third, although Eq. (37) indicates three scaling subranges with scaling exponents being -17/3, -13/3 and -3, the contribution of -13/3 scalar spectrum is too small that has been covered by the -17/3 scalar spectrum. Finally, when $c_u$ and $c_s$ are increased with a fixed $Sc = c_u/c_s = 0.1$, both the -17/3 and -3 scalar spectra move towards low wavenumber region.

**6. Discussion**

To this end, we have attempted to establish a comprehensive picture for velocity and scalar cascades in momentum-scalar coupling turbulence, regarding anomalous diffusions of momentum and scalar, which determine how large wavenumber the turbulent kinetic energy and scalar variance can be delivered into. For the cases that $\beta < 2/3$ or without forcing, the inertial subrange for either kinetic energy or scalar is directly intersected with the dissipation subrange. The results are compatible for conventional hydrodynamic turbulence, passive scalar turbulence and buoyancy-driven turbulence where momentum and scalar transports are coupled.

On the other hand, several limitations should be acknowledged as well. First, the analysis assumes homogeneity, isotropy, and the parallel alignment of $\boldsymbol{M}$ and $\boldsymbol{N}$. Deviations from these conditions would introduce anisotropic corrections and additional coupling terms. Second, the marginal cases $\gamma = 2/3$ and $\alpha = 2/3$, where the convective and diffusive time scales share the same $k$-dependence, are excluded from the present analysis. Third, some predicted characteristic wavenumbers may exceed the physical bounds of the continuum hypothesis, particularly for small $\alpha$, where the predicted wavenumbers can approach or surpass the wavenumber corresponding to Planck length where quantum gravity become dominant. This suggests either that such extreme anomalous exponents are physically unrealizable in real fluids (needs a cut-off), or additional physical mechanisms (e.g., quantum effects or molecular cutoffs) become relevant before such wavenumbers are reached. The predictions should therefore be interpreted as asymptotic scaling laws valid within the regime where the fractional diffusion description remains physically meaningful. Part 1 has focused exclusively on long-range forcing ($\beta < 2/3$) or unforced turbulence, where the forward cascade picture is retained. A further discussion on short-range forcing ($\beta > 2/3$) and its application to electrokinetic turbulence will be elucidated in a sequential manuscript (Part 2).

A notable by-product of the present analysis is a quantitative explanation of the bottleneck effect in the compensated kinetic energy spectrum. The terms $k^{-1/3}$ and $k$ in Eq. (25), arising from the linearized flux–dissipation feedback, produce a spectral bump whose locations ($k/k_K \approx 0.040 \sim 0.153$) are consistent with both experimental [35] and numerical [34] observations. This suggests that the bottleneck is not merely a hyperviscous artifact—as previously debated in the context of Galerkin-truncated simulations [38-40]—but emerges naturally from the scale-dependent imbalance between the energy flux and its dissipation rate. The present model captures this feature even without hyperviscosity, indicating that the bottleneck originates from the nonlinear flux–dissipation interplay rather than from numerical or artificial dissipation mechanisms.

Table 1 summarizes the subranges of scalar turbulence, their conditions, scalar spectra, and the counterparts in classical models for comparison. Additional fine scales and scaling exponents in scalar turbulence can be concluded from the diversely anomalous diffusions of both momentum and scalar, regarding $\gamma$, $\alpha$, $Sc_Z$, $\varepsilon_u$ and $\varepsilon_s$ .

Table 1. Scalar cascade regimes in the homogeneous and isotropic turbulence without external force and with long-range external force of $\beta < 2/3$.

<table>
<tr><th>Regime</th><th>Conditions</th><th>Scalar spectra</th><th>Classical counterparts</th></tr>
<tr><td>Inertial-convective</td><td>$Sc_Z \gg 1$</td><td>$E_s \sim k^{-\frac{5}{3}} \exp\left[\frac{2\mathrm{Re}\left(e^{i\frac{1}{2}\pi\alpha}\right)}{\left(\alpha - \frac{2}{3}\right)\sqrt{C_K}}\left(\frac{k}{k_S}\right)^{\alpha - \frac{2}{3}}\right]$</td><td>Obukhov–Corrsin (1949)</td></tr>
<tr><td rowspan="2">Viscous-convective</td><td>$Sc_Z \gg 1$</td><td rowspan="2">$E_s \sim k^{-1} \exp\left[2\mathrm{Re}\left(e^{i\frac{1}{2}\pi\alpha}\right)\alpha^{-1}\left(\frac{k}{k_B}\right)^{\alpha}\right]$</td><td>Batchelor (1959)</td></tr>
<tr><td>$Sc_Z \ll 1, \gamma > \alpha$</td><td>No classical counterpart</td></tr>
<tr><td>Inertial-diffusive</td><td>$Sc_Z \ll 1, \gamma \le \alpha$</td><td>$E_s \sim \left(\frac{k}{k_{S3}}\right)^{-\frac{5}{3}-2\alpha} + \left(\frac{k}{k_{S4}}\right)^{\gamma-2\alpha-\frac{7}{3}} + \left(\frac{k}{k_{S5}}\right)^{2\gamma-2\alpha-3}$</td><td>Batchelor–Howells–Townsend (1959)</td></tr>
</table>

## 7. Conclusions

To this end, we have developed a generalized theoretical model for momentum–scalar coupled turbulence accounting for anomalous diffusion of both fields, described by fractional biharmonic operators. For long-range forcing ($\beta < 2/3$) or unforced turbulence, we derived analytical expressions for the kinetic energy spectrum and scalar spectrum, and identified the characteristic dissipation wavenumbers $k_K = \left(\varepsilon_u^{1/3}/c_u\right)^{1/\left(\gamma-\frac{2}{3}\right)}$ (reciprocal of Kolmogorov scale) and $k_S = \left(\varepsilon_u^{1/3}/c_s\right)^{1/\left(\alpha-\frac{2}{3}\right)}$ (reciprocal of scalar dissipation scale). The well-known bottleneck effect is explained by a linear approximation of dissipation on kinetic energy flux. The theory also demonstrates how the anomalous diffusion of momentum (i.e. $\gamma$) affects bottleneck effect, showing consistency with previous investigations on hyperviscosity. Counter-intuitively, superdiffusion inhibits diffusion at high wavenumber and broadens the inertial range by increasing both $k_K$ and $k_S$.

The scalar spectra in different scaling subranges are also investigated. To be compatible with anomalous diffusions of momentum and scalar, the classical Schmidt number is replaced by a scale-dependent anomalous Schmidt number $Sc_Z = k_0^{\gamma-\alpha} c_u/c_s$, which governs the cascade topology. For $Sc_Z \gg 1$, we recover a generalized viscous-convective subrange with $E_s \sim k^{-1}$. This result can also be realized for $Sc_Z \ll 1$ when $\gamma > \alpha$. While for $Sc_Z \ll 1$ and $\gamma \leq \alpha$, an inertial-diffusive subrange according to Batchelor, Howells and Townsend [45] can be reproduced, with $E_s \sim k^{-17/3}$ at $\gamma = \alpha = 2$. From the current model, all classical scalings are recovered when $\gamma = \alpha = 2$. Therefore, the framework provides testable predictions for a wide class of turbulent systems where non-Fickian transport dominates the small-scale dynamics.

**Acknowledgement** The investigation is supported by the Open Project of the Shaanxi Provincial Key Laboratory of Optoelectronic Technology (SXPEL2026O-02).